\RequirePackage{fix-cm}
\documentclass[smallextended]{svjour3}       % onecolumn (second format)
\smartqed  % flush right qed marks, e.g. at end of proof
\usepackage{graphicx}
\usepackage{rotating}
\usepackage{multirow}
\usepackage{color}

\newcommand{\CB}[1]{{\color{black}{#1}}} 

\begin{document}

\title{Cryptocurrency market structure: connecting emotions and economics}


\author{Tomaso Aste}


\institute{  Department of Computer Science, UCL, London, UK. \and 
UCL Centre for Blockchain Technologies, UCL, London, UK. \and
Systemic Risk Centre, London School of Economics, London UK.
           \email{t.aste@ucl.ac.uk}         }

\date{\today}

\maketitle

%
%{\it For submission, the full paper, a blind copy and a separate cover page should be sent to the special issue ed- itorial board via email at the following (clickable) ad- dress: abarletta@econ.au.dk. All submissions should be in PDF. The cover page should include: the title; the full names of all authors (first and last); the authors? institutional affiliations; the name, address, telephone number and e-mail address of the author responsible for 
%
%Deadline 15/09/18 !!!!
%
%https://www.springer.com/finance/journal/42521
%}
\vskip1.cm

\begin{abstract}
We study the dependency and causality structure of the cryptocurrency market investigating collective movements of both  prices and social sentiment related to almost two thousand cryptocurrencies traded during the first six months of 2018.
\CB{
This  is the first study of the whole cryptocurrency market  structure. It introduces several rigorous innovative methodologies applicable to this and to several other complex systems where a large number of variables interact in a non-linear way, which is a distinctive feature of the digital economy.
The analysis of the dependency structure reveals that prices are significantly correlated with sentiment. 
The major, most capitalised cryptocurrencies, such as bitcoin, have a central role in the price correlation network but only a marginal role in the sentiment network and in the network describing the interactions between the two. 
The study of the causality structure reveals a causality network that is consistently related with the correlation structures and shows that both prices cause sentiment and sentiment cause prices across currencies with the latter being stronger in size but smaller in number of significative interactions.   
Overall our study  uncovers a complex and rich structure of interrelations where prices and sentiment influence each other both instantaneously and with lead-lag causal relations. 
A major finding is that  minor currencies, with small capitalisation, play a crucial role in shaping the overall dependency and causality structure.
Despite the high level of noise and the short time-series we verified that these networks are significant with all links statistically validated and with a  structural organisation consistently reproduced across all networks.
}
\end{abstract}
\section{Introduction} 
During the last two years we have witnessed the creation of a large number of cryptocurrencies.  
This burst has been mainly fueled by the opportunity generated by the ICO mechanism used by companies as a new channel to fund innovation. Furthermore, this burst follows the surge of new business models based on blockchain and associated digital tokens and crypto-money. 
\CB{The most dynamic period in the cryptocurrencies market has been, so far, the beginning of 2018 on which this study is focusing.}
At the time of writing (September 2018) the cryptocurrency market capitalization is floating around 200 billion USD down from 800 billion USD reached in January 2018 \cite{marketcap}.
This market comprises thousands of currencies with only a few with significant capitalization. In particular five currencies, namely, Bitcoin (BTC), Bitcoin Cash (BCH), Ethereum (ETH), Litecoin (LTC) and Ripple (XTC) have been dominating the market during the last few years with a share of capitalization consistently above 70\%. 
Overall, there are 15 currencies with capitalization over 1 billion USD, more than 60 with capitalization over 100 million USD and about 800 with capitalization over 1 million USD.
This is a new and confused market characterized by large volatilities, by quick increases in the value of some currencies at the time of their release and, often, a rapid decrease of the value afterwards until failure. 
This is a market strongly echoed in social media with great expectations, quick swifts of sentiment, strong beliefs and harsh  disputes.

In the literature, there have been some studies of correlations in cryptocurrency markets highlighting the non-normal statistics of correlations between price fluctuations \cite{gkillas2018extreme} and their relations with fiat currencies \cite{szetela2016dependency}. 
Social media and Twitter sentiment signals have been used to attempt nowcasting and forecasting for some of these currencies \cite{kim2016predicting,kaminski2014nowcasting}. The main focus, so far, has been on Bitcoin with little published research on other cryptocurrencies. 

In this paper we investigate how cryptocurrency prices collectively behave 
\CB{ and how the price behaviour is related with the sentiment behaviour expressed through Twitter and StockTwits \cite{stocktwits} messages that refer explicitly to the related currency.} 
We ask if this market has a characteristic structure, we enquire where the major cryptocurrencies are located within this structure \CB{and we investigate the role of minor cryptocurrencies in shaping this structure}. We study the influence of social sentiment and its interplay with prices. We do this by looking at the entire market (1944 cryptocurrencies recorded during the first six months of 2018) instead of concentrating on a few `important' currencies only.
\CB{We intentionally study the whole market even if most of the capitalization is retained by a few currencies and most of the other  currencies play a marginal economic role. 
%Indeed, these minor currencies might be insignificant from an economic perspective,  however, in this work we uncover signals revealing that  these marginal currencies  play a statistically significant role in the collective dynamics of prices and their interplay with social sentiment. 
From a naive perspective, a-priori one would had expected to observe minor currencies being driven by the behaviour of the major ones in a similar way as it happens for the dynamics of stock prices that tend to cluster around the leading firms of the relative sector \cite{aste2010correlation,song2012hierarchical,musmeci2014clustering}.  
Surprisingly,  we shall uncover instead that this is not happening in the cryptocurrency market. Indeed, in this work we uncover signals revealing that  these marginal currencies  play a statistically significant role in the collective dynamics of prices and their interplay with social sentiment. 
Therefore they should not be excluded a-priori from the investigation and their role with respect the major  currencies must be studied in detail.
This opens new challenges for what concerns investment strategies and risk management which must handle very large number of variables and cannot be limited to the study of a few influential factors. 
}

In this market, both prices and sentiment data are noisy with large volatility; for this reason we quantify dependency and causality mainly using rank statistics and topology reducing in this way the effect of noisy outliers. 
\CB{We pay a special attention to statistically validate dependency and causality links by using non-parametric permutation tests and by assessing the effect of the validation threshold on the resulting structure. 
We also cross-test results by comparing the overall structural properties of the networks discarding the null-hypotersis that they might be the expression of random spurious links.}
Our study uncovers a complex structure of interrelations where prices and sentiments influence each other both within a given currency and across currencies. 
To our knowledge, this is the first attempt to understand dependency and causality structure in this market. 

\CB{The structure of the cryptocurrency market as unveiled in this work is unavoidably specific to the period investigated, which has been a very special and dramatic period. In this respect, this paper presents a unique picture of a very interesting period of the cryptocurrency market. Despite the fact that already at the time of finishing the revision of this paper the cryptocurrency market has changed significantly, nonetheless some aspects such as the intrinsic nonlinearity in the interactions and the role of `minor' variables on the whole system will rest significant for this market as well as for other systems in the digital economy. Furthermore, this paper contributes to the study of these systems by introducing several general and rigorous methodologies to handle dependency and causality in these noisy and non linear systems composed by a large number of variables and often supported by a small number of observations. These novel methodologies have broad applicability to the study of the digital economy and complex systems in general.

The paper is organised as follows. In section \ref{S.2} we describe the dataset. Section \ref{sec:related:work} describes the methodology adopted for quantifying dependency, causality, their representation into networks and the statistical validation procedure. Results are presented in Section \ref{S.results} where the properties of dependency and causality networks for both sentiment and prices and their interplay are described in details. Section \ref{S.discussion} provides a detailed discussion of the results with special attention at their statistical significance.
Conclusions and perspectives are outlined in Section \ref{S.conclusion}.
}

\section{Data} \label{S.2}

\begin{figure}
\centering
\includegraphics[width=.95\linewidth]{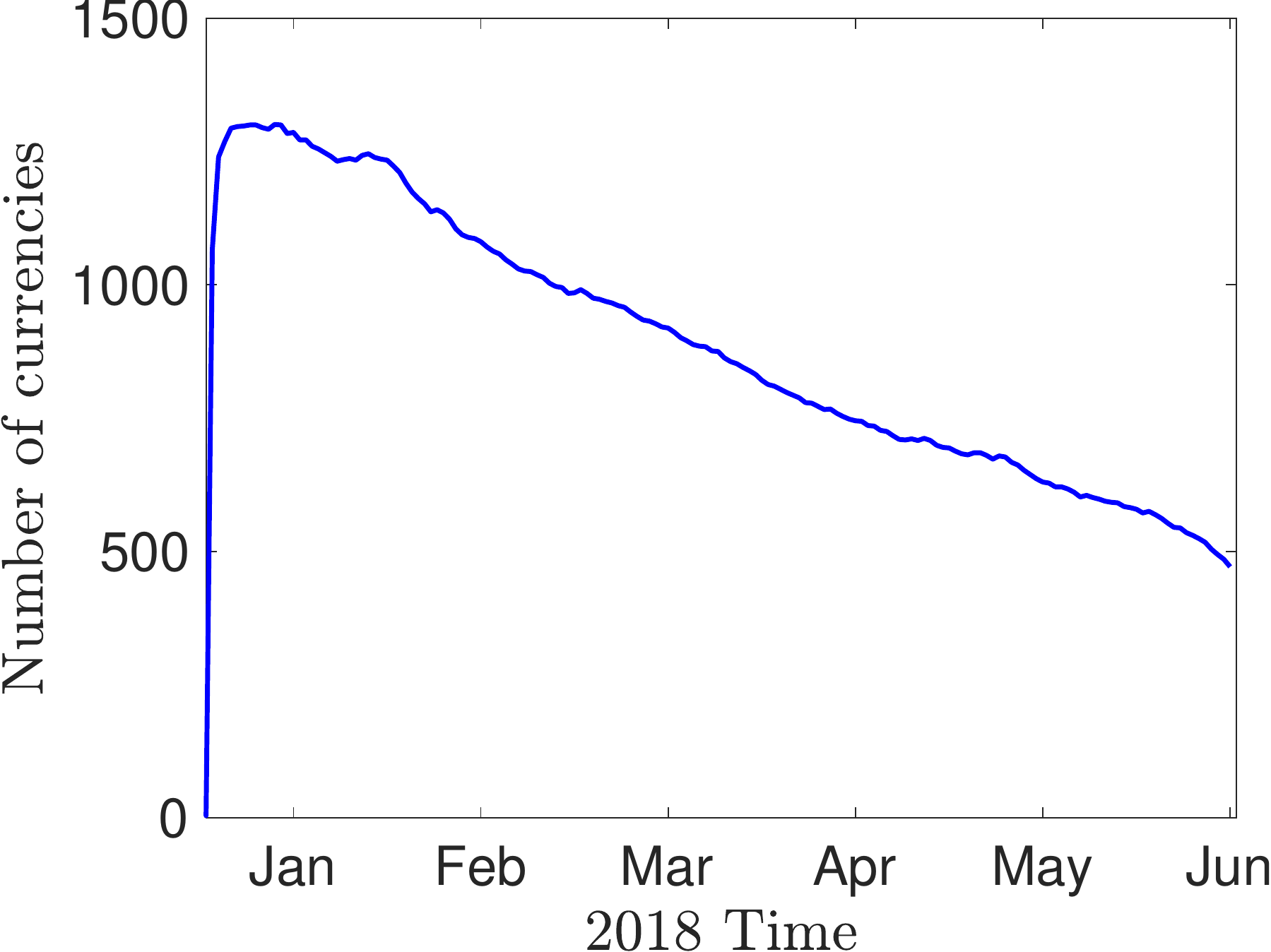}
\caption{Number of currencies simultaneously present during the period Jan-Jun 2018.}
\label{Fig_numCurr}
\end{figure}

Prices and Twitter sentiment data of 1944 cryptocurrencies traded during the period from  January 2018 (02/01/2018) to  the middle of of June (14/06/2018) are analyzed. 
In the dataset,  four major  currencies, namely BTC,  LTC, ETH and XRP  had records starting earlier, respectively from:  01/09/2014, 01/09/2014, 07/08/2015 and 21/01/2015.
The number of currencies simultaneously present at any time during the period Jan-June 2018 is reported in Fig.\ref{Fig_numCurr}. 
\CB{This number is not constant because new currencies are introduced over time and other fail and cease to be traded in the market. Often they do not disappear but their capitalisation become negligible and the price become constant and they are therefore excluded from our database.}
The largest number of currencies contemporarily present were 1301 as recorded at the  end of January 2018. Then numbers gradually decreased to 471 at the end of the observation period. 
\CB{The peak at the  end of  January 2018 reflects the popularity of ICOs that indeed peaked in that period.}
Prices have been obtained from Cryptocompare \cite{cryptocompare} whereas sentiment is provided by PsychSignal \cite{PsychSignal}.
The sentiment signal is computed from natural language processing of Twitter and StockTwits \cite{stocktwits} messages that refer explicitly to the related currency. \CB{Messages are classified as positive, negative or unclassified depending on the words contained and their context. 
The signal we analyse is the number of messages in each category, referred to as volume.}
\CB{In this work we consider the relative changes in positive and negative volumes only; we treat them as separate signals and we ignore the unclassified volumes.}
Original data are hourly, though in the following analytics we transformed them into a daily signal by aggregating prices  reporting the average daily price and by aggregating volumes  reporting the total daily volume. This aggregation process reduces noise. Similar results are obtained with different aggregation criteria. 

 % https://www.cryptocompare.com/

\section{Methodology} \label{sec:related:work} 

We investigated collective movements of currency prices and currency sentiment by computing  Kendall cross-correlations \cite{kendall1938new} and non-parametric transfer entropy \cite{schreiber2000measuring,tungsong2017relation} of daily log-returns, \CB{ $\log Price(t)- \log Price(t-1)$ }(differences of the logarithm of the price between a day and the previous), and daily changes of the logarithm of the number of messages classified  positive or negative, $\log$({\it Number of messages with positive sentiment on day} $t$)$-\log$({\it Number of messages with positive sentiment on day} $t-1$). 
The choice of the log-returns for prices is standard in financial literature \CB{ \cite{campbell1997econometrics}.} Differencing makes the series stationary and the logarithm reduces effects of non-normal variations. 
In contrast, the choice of log variation of sentiment volume is here mainly motivated by the convenience of treating both variables in the same way. Test results show that the use of the volume-variations instead of its log-variations gives overall similar outcomes.  

We estimated dependency structure by computing Kendall's $\tau$ correlation coefficients \cite{kendall1938new}.  
\CB{We verified that comparable results are obtained by using Pearson or Spearman correlations. 
Nonetheless, Kendall correlation are a more appropriate analytics tool for the kind of data we are investigating in this work. 
Indeed, the} time-series are short and  the  statistics of both sentiment and prices log-variations are non-normal, making a rank estimate more reliable to establish dependency than the Pearson's counterpart \cite{kendall1938new,pozzi2008dynamical,pozzi2008centrality}.

Correlations were  computed between pairs of variables by using all available days where both variables had  observations. 
We consider only correlations between pairs of variables with more than 20 common observations.
\CB{We validated non-parametrically correlations by using a permutation test that compares the observed correlation coefficients with a null (non-correlated) hypothesis generated by randomly shuffling time entries in the series. }
Observed correlations are considered `valid' only if they deviate from the mean of the random ones by at least three standard deviations (i.e. $Z$ score larger than 3 \cite{wilks1932certain}). 
\CB{Note that this validation criteria is non-parametric and therefore robust also in the present case where correlations do not follow the ststistical distribution assumed in standard tests  \cite{kendall1946advanced}.}
%We also repeated all measures for validation with $Z>6$. Results were consistent.  

The dependency structure was analyzed in terms of its topological properties (the validated links structure). 
For this purpose, we define the network's adjacency matrix $A_{i,j}$ as a matrix with $A_{i,j}=1$ when the corresponding correlation has $Z>3$ and it is computed from more than 20 observations; $A_{i,j}=0$ otherwise. 

We computed all combinations of correlations within and across the vaiables: i) cross correlations of log-price returns; ii) cross correlations of  log-volume sentiment changes \CB{(for both positive and negative sentiment)}; iii) the combined cross correlations between price and sentiment log changes \CB{(for positive sentiment only)}.

\CB{We also investigated weighted betweenness-centrality and closeness measures   \cite{newman2008mathematics}  for each node in the validated correlation networks. 
The weight of an edge $(i,j)$ between currency `$i$' and currency `$j$'  was associated to the relative correlation $\tau_{i,j}$ as $w_{i,j} = 1-\tau_{i,j}^2$.}
Therefore uncorrelated nodes are connected with edges with cost equal to 1 and perfectly correlated or anti-correlated nodes have zero-cost connection. 

%Causality is studied by computing transfer entropy \cite{TE}. ....
%A validated transfer entropy network is computed by .... (Z>3 and 40 common observations)...
Causality was studied by estimating transfer entropy computed by means of a non-parametric histogram methodology, using 4 equally spaced bins (see in \cite{tungsong2017relation}).
\CB{ Transfer entropies were computed for  log-price returns and log-volume positive sentiment changes.
A validated transfer entropy network was constructed in an analogous way to the validated correlation networks by keeping links generated from time-series combinations longer than 40 days and keeping transfer entropy permutation-test $Z$ score larger than 3. }
Transfer entropy measures the reduction in uncertainty about the value of a given variable provided by the knowledge of the previous values of another variable discounting for the information from the past of the variable itself.
\CB{In our case, we tested the causal effect of  positive sentiment on the next day prices and -conversely- the causal effect of prices on next day positive sentiment across all currencies.}
\CB{
We also compared transfer entropy results with the Granger causality approach that uses linear regression \cite{granger:econ,granger1980testing}. The outcomes of the two methods are overall consistent and here we report only the results for the non parametric method that obtains a larger number of validated causal links.
It must be noted that, in the linear case, when variables follow a multivariate normal distribution, the transfer entropy method is identical to the well-known Granger causality approach \cite{PhysRevLett.103.238701}. 
However, we are well aware that the dataset under investigation is not following a multivariate normal distribution and therefore the non-parametric transfer entropy approach must be adopted. 
The fact that we obtain a larger  number of valid links with  non-parametric transfer entropy reinforce the point that this system of variables must be properly described with non-normal multivariate statistics.
}
For the histogram approach we tested different binning observing that  results are affected by the choice of the  bins but overall outcomes are consistent over a range of bins from 3 to 6.

Under normality assumptions a $Z$ score larger than 3 would imply rejection of null hypothesis with p-value below 0.13\%. 
In this paper we use $Z>3$ as a threshold to eliminate noise from the correlations and we do not directly associate this threshold on the $Z$-score  with p-value null hypothesis rejection.
Indeed, in our case, p-value is affected by the fact that statistics are not normal and samples are small. 
%There are potential corrections, for instance by using Student-t distribution instead of normal and by scaling  p-value with Bonferroni correction.
A precise testing of statistical significance is beyond the purposes of this paper however it is crucial to establish if the structures that we uncover are reflecting dependency and causalities among the variables or they are just picking randomly spurious interactions from a large number of possibilities on very noisy data.
To this purpose we also tested validation at $Z>6$ which, under  normality assumptions, would imply rejection of null hypotheses with p-value below $10^{-9}$. 
Outcomes from $Z>6$ were consistent with the analysis with $Z>3$ but networks become extremely sparse to the point that the transfer entropy network becomes largely disconnected into small clusters and isolated nodes. 
We therefore also looked at similarity between the various networks using the network from cross-correlation of log-price returns as a structure-template.  The hypothesis we tested in this case was that significant structural similarity being incompatible with random networks.

\section{Results}  \label{S.results}
%Here we report results for the structure of the correlation networks the validated correlation network. 
%Isolated nodes are not acounted in the degree distributions. 

\subsection{ Price-Price \& Sentiment-Sentiment cross-correlation validated networks }

\begin{figure}
\centering
\includegraphics[width=.45\linewidth]{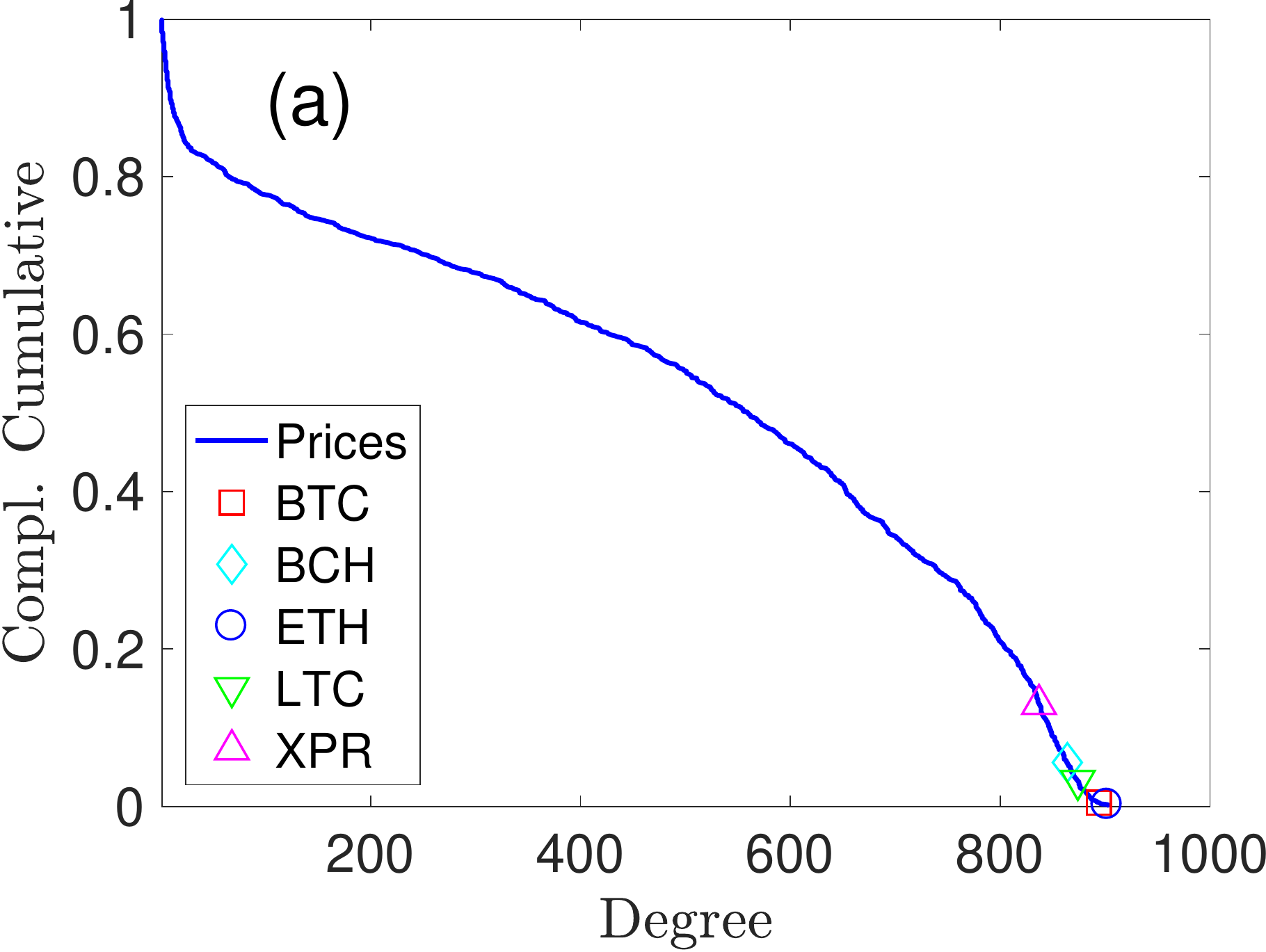}
\includegraphics[width=.45\linewidth]{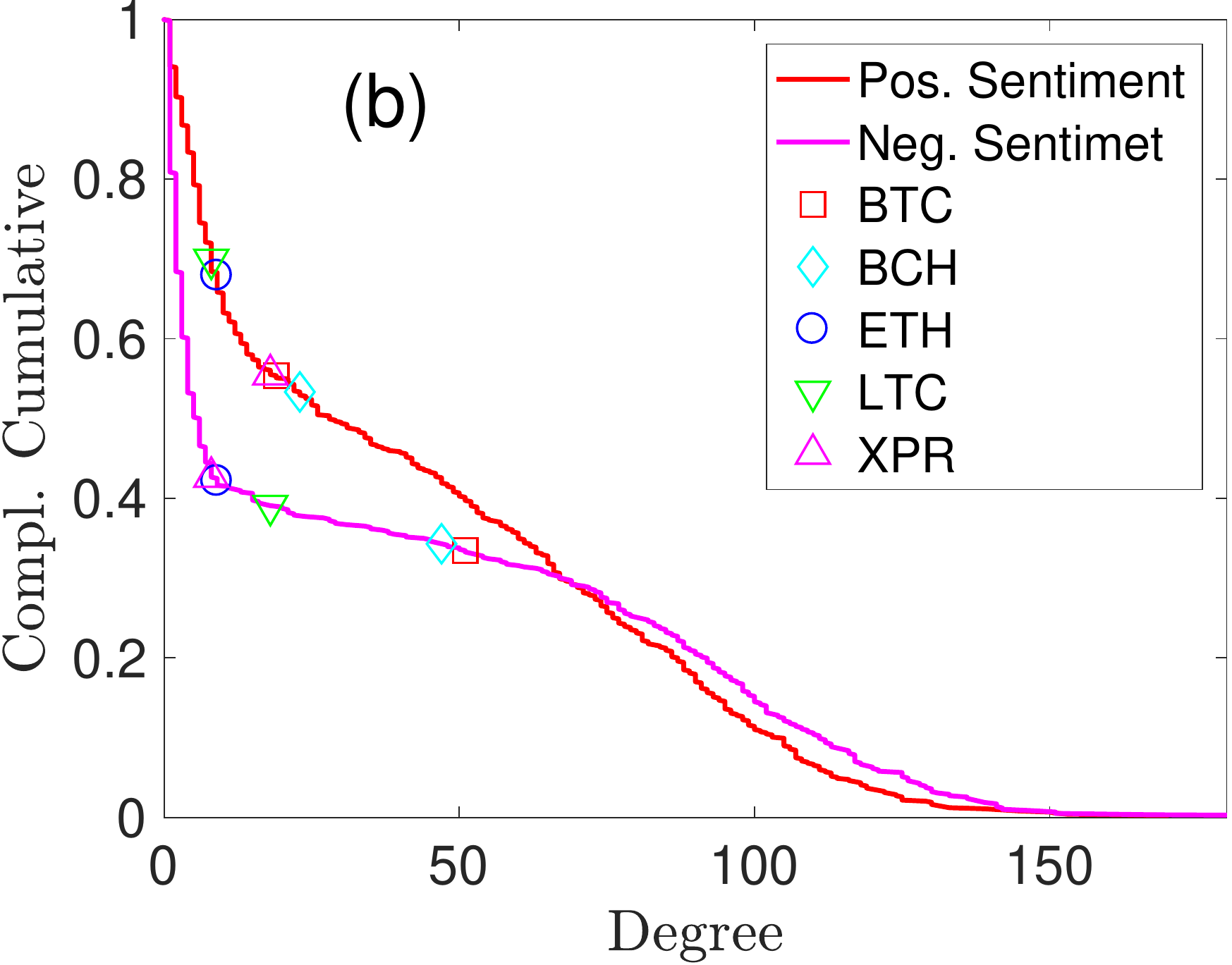}
\caption{Complementary cumulative degree distribution ($Probability(k > x$) for the validated Kendall cross correlation networks constructed from (a) the cross correlations of  log-price returns  and (b) cross correlations of  log-volume sentiment changes for both positive and negative sentiments.
The degrees of Bitcoin (BTC), Bitcoin Cash (BCH), Ethereum (ETH), Litecoin (LTC) and Ripple (XTC) are indicated explicitly with symbols.}
\label{Fig_xPrices}
\end{figure}

\begin{figure}
\centering
\includegraphics[width=.45\linewidth]{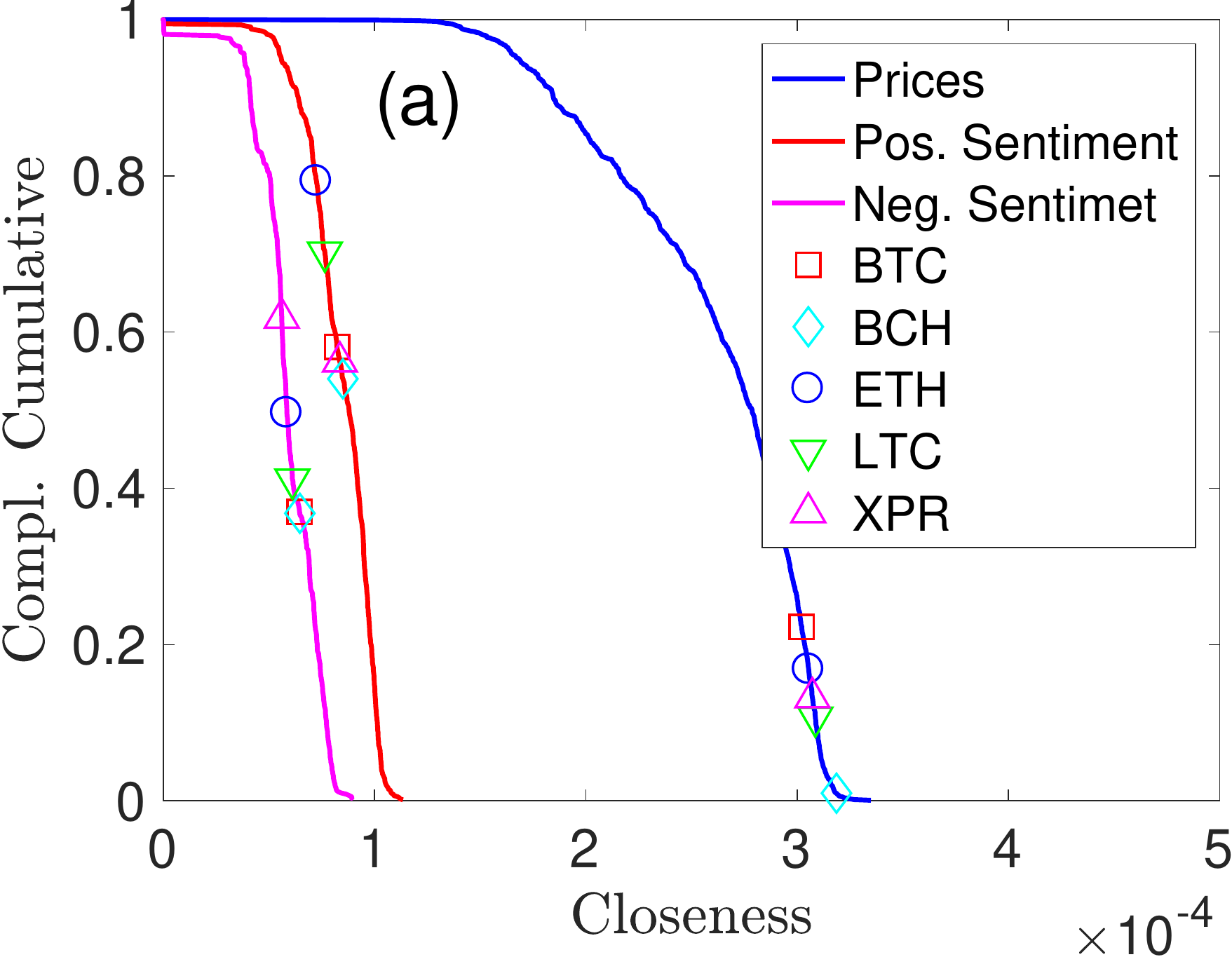}
\includegraphics[width=.45\linewidth]{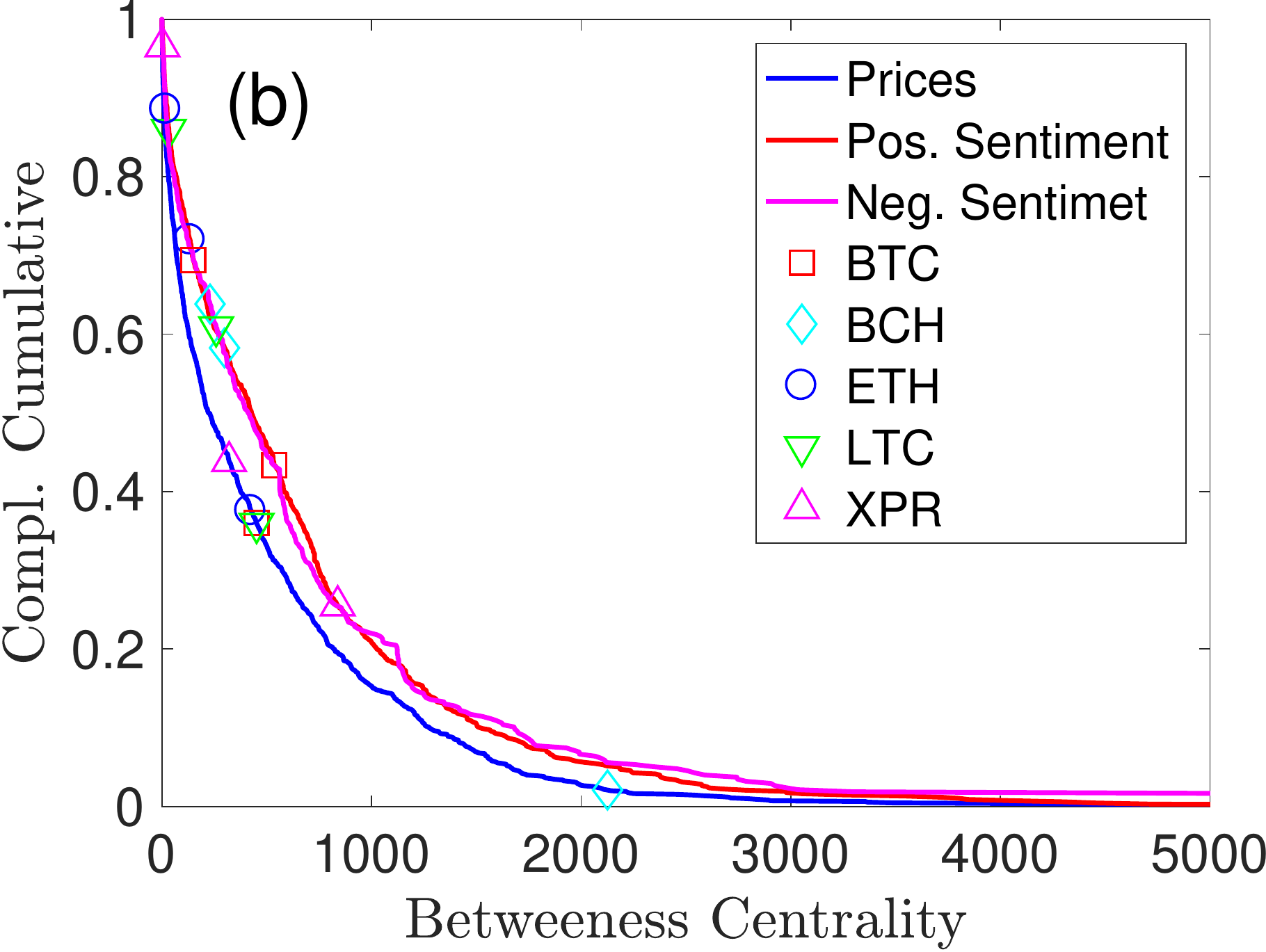}
\caption{Closeness and betweenness-centrality complementary cumulative probability distributions computed over the validated networks using weights $w_{i,j} = 1-\tau_{i,j}^2$.}
\label{Fig_ClosBetw}
\end{figure}

We first computed the validated networks from cross correlation of: 1) log-prices; 2) positive sentiment log-volume variations;  3) negative sentiment log-volume variations. 
These are symmetric matrices  of size $1944\times1944$ with ones on the diagonal.
We observed predominately positive correlations with average correlation between log-prices variations being equal to 0.40, average correlation between positive sentiment log-volume variations being equal to  0.18 and average correlation between the negative sentiment log-volume variations being equal to 0.22.

We computed the degree distribution by considering for each currency $i$ the number of other currencies $j$ with which it shares a statistically validated correlation ($k_i=\sum_j A_{i,j}$).
The valid correlation networks are sparse with the network from price log-returns correlations having 15\% of valid links and average degree of 300.7. In contrast, the positive and the negative sentiment volume networks have respectively average degrees equal to 16.3 and 10.7. 
All networks have one connected giant component, a few small clusters and several isolated nodes. The sizes of the giant components are respectively 1216, 730 and 564 for price, positive and negative sentiment networks.
Results for the complementary cumulative  degree distributions (Probability($k_i > x$)) are reported in Figs.\ref{Fig_xPrices}(a,b) for the three networks. 
In the figures the degrees of Bitcoin (BTC), Bitcoin Cash (BCH), Ethereum (ETH), Litecoin (LTC) and Ripple (XTC) are indicated with symbols. 
A summary of the results for the major currencies is reported in Table \ref{Tab_degrees}.
%We notices the very high degrees in the price valid cross-correlation network and the larger degrees in the negative sentiment network with respect to the positive one.
We notice that  in the price network these major cryptocurrencies have high degrees between 800 and 900 ranking in the top 10\% of highly connected nodes being therefore hubs within the connected component.
Conversely, these currencies have relatively low degrees in the sentiment networks ranking below 50\% in the positive sentiment network and just above 50\% in the negative sentiment network with number of connections between 10 and 50.

In order to better understand the relative positioning within the cryptocurrency  market also with respect to the weighting of the correlations, we computed closeness and centrality distributions. These weighted measures, computed over the validated networks,  are reported in Fig.\ref{Fig_ClosBetw}.
We observe that for the closeness the relative ranking of the  five major  cryptocurrencies is similar to the ones observed for the degree distribution; conversely the betweenness-centrality places all major cryptocurrencies into medium/peripheral rankings.

\subsection{ Price-Sentiment validated correlation network  }
\CB{From now on we consider only positive volume sentiment. This  choice is to simplify computation and description of the results.}
We  investigated the Kendall cross correlations between  log variation of \CB{positive sentiment }volume and log variations of price. 
This is an asymmetric $1944\times1944$ matrix representing a bipartite undirected network.

The diagonal elements of this  matrix are the correlations between positive sentiment and price for each currency. 
 Among the  five major  cryptocurrencies we observe correlations on the diagonal of:   0.09 BTC,  0.07 BCH, 0.11 ETH,  0.10 LTC and  0.05 XPR. Except for BCH and XPR they are all statistically validated with $Z> 3$ and series length over 20 points (BCH and XPR have instead $Z=1.1$ and  $1.7$ respectively).
Overall, only 1\% of  currency log-price variations have a valid correlation with their own log positive sentiment volume variations; they have mostly positive correlations but there are a few with negative valid correlations as well. 

\begin{figure}
\centering
\includegraphics[width=.95\linewidth]{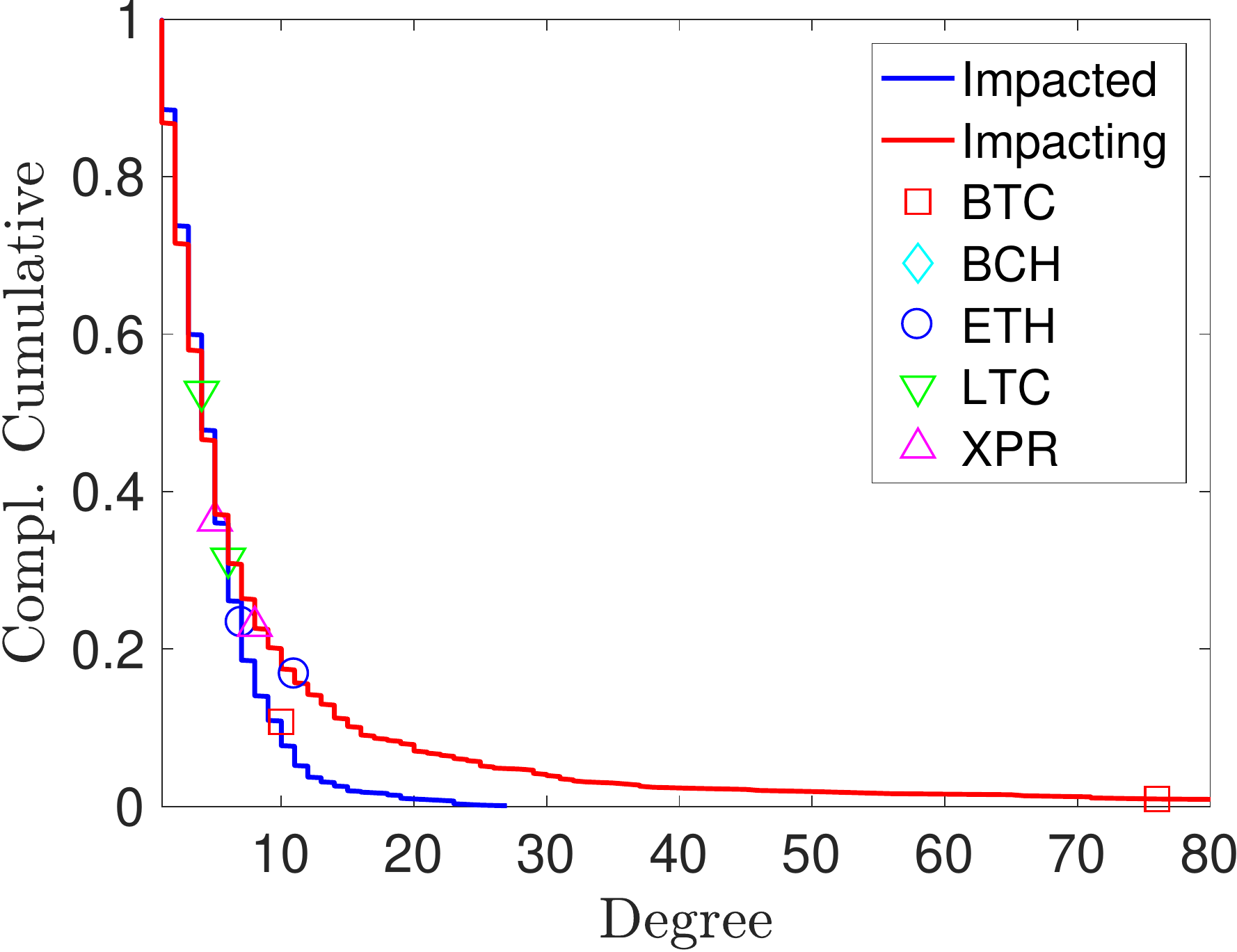}
\caption{In-degree and out-degree complementary cumulative  distributions for the validated Kendall cross correlation network between log variations of price of one currency and log variation of positive sentiment volume of another. 
The `impacted' distribution is counting the number of valid links with other currencies whose positive sentiment  is affected by the currency price.
The `impacting' distribution is counting the number of valid links with other currencies whose price is  affected by the currency positive sentiment.}
\label{Fig_SentimetPrice}
\end{figure}

The off-diagonal elements, $\tau_{i,j}$ $i\not=j$, of this matrix  are non-symmetric ( $\tau_{i,j}\not=\tau_{j,i}$). 
They represent respectively: $\tau_{i,j}$ the correlation of positive sentiment of currency $i$ with price  of currency $j$;  $\tau_{j,i}$  the correlation of positive sentiment of currency $j$ with price of currency $i$.
Here we must distinguish two kinds of degrees:
 1)  {\bf `impacting' degree} which is the sum of the valid entries over the columns ($Ig_i=\sum_{j}A_{i,j}$); 
 2) {\bf `impacted' degree} which is the sum of the valid elements over the rows ($Id_j=\sum_{i}A_{i,j}$).
 Note that, in the literature, these degrees are commonly referred as in-degree and out-degree \cite{newman2008mathematics}; however in our case this underlying implicit representation of the graph as a directed graph can be misleading \CB{implying some sort of causality that is not measured here (it will be measured with Transfer Entropy as reported in the next session)}.  
The  `impacting' degree of a given currency $i$ is counting the number of valid links with other currencies $j$ whose price is  affected by the currency positive sentiment. 
Conversely   `impacted' degree of a given currency $i$ is counting the number of valid links with other currencies $j$ whose sentiment is  affected by the currency price. 
It results that this off-diagonal matrix has 0.2\% validated entries.  
The average degree is  3.1 for both impacting and impacted degrees.
The degree distributions are reported in Fig.\ref{Fig_SentimetPrice}.
We observe that the distribution of the impacting degree has fatter tails than the one of the  impacted degree indicating that large variations of sentiment of a given currency are more influential on other currency price variations than large changes in currency price to other currency sentiment.
Given that the average degree is the same for both distributions we have -conversely- that  small variations of sentiment of a given currency is more influential to other currency prices variations than small changes in currency price to other currency sentiment.
In particular we observe that changes in Bitcoin sentiment are correlated above validation threshold with changes in prices of almost eighty other currencies whereas  changes in Bitcoin price have valid correlation links to only ten other currency sentiment changes. \CB{A summary of the results for the major currencies is reported in left columns of Table \ref{Tab_degrees}.}

We must stress that correlation is not causality and from the previous results we cannot conclude what is the cause and what is the effect. For this purpose we must use other kinds of measures, such as -for instance- transfer entropy, as we shall proceed to the next section.

\begin{table}
\begin{center}
\begin{tabular}{ |c|c||c|c|c||c|c||c|c|c|c| } 
\hline
% & Corr & Corr   & Corr & Corr &  Corr & TE  &  TE  &  TE  & TE \\
% &  &    &    & Ig &   Id &   Ig &   Id &   Ig &   Id \\
%  & P-P & pS-pS  & nS-nS  & pS-P & pS-P & P$\rightarrow$pS  &  P$\rightarrow$pS  &  pS$\rightarrow$P  &  pS$\rightarrow$P  \\
 &
 &\begin{turn}{-90} Price valid cross correlation degree \end{turn} 
 &\begin{turn}{-90}  Positive sentiment valid cross correlation degree \end{turn}  
 &\begin{turn}{-90}  Negative sentiment valid cross correlation degree \end{turn}  
 &\begin{turn}{-90}  Positive sentiment impacting other currencies prices \end{turn}  
&\begin{turn}{-90}  Price impacted by other currencies positive sentiment \end{turn}  
&\begin{turn}{-90}  Price causing other currencies positive sentiment \end{turn}  
&\begin{turn}{-90}  Other currencies prices causing positive sentiment \end{turn}  
&\begin{turn}{-90}  Positive sentiment causing other currencies prices \end{turn}  
&\begin{turn}{-90}  Other currencies positive sentiment  causing price \end{turn}  
 \\
\hline
\hline
 \multirow{ 5}{*}{$Z>3$} 
 & BTC & 894& 19& 51&76 & 10&11 &10 &15 &8\\  \cline{2-11}%\cline{2-11}
& BCH & 864& 23& 47& 3& 3&2 &4&13 &13 \\  \cline{2-11}
& ETH & 902& 9& 9&11 & 7& 7 &27&6 &8 \\  \cline{2-11}
& LTC & 874& 8& 18&4 & 6&10 &17&17 &22 \\  \cline{2-11}
& XPR & 837 & 18& 8&8 & 5 &2 &16 &6 &11 \\ 
 \hline
 \hline
\multirow{ 5}{*}{$Z>6$} 
& BTC & 542& 2 & 2  &1  & 1  &0 &0 &1 &0\\  \cline{2-11}
& BCH & 497& 1 & 2 & 0  & 0  &0 &0&0 &0 \\  \cline{2-11}
& ETH & 535& 1 & 2  &0  & 0  &0 &0&0 &0 \\  \cline{2-11}
& LTC & 484& 1 & 1  &0  & 0  &0 &1 &0 &0 \\  \cline{2-11}
& XPR & 507&0 & 1  &0  & 0  &0 &0 &0 &0 \\ \hline
\end{tabular}
\end{center}
\caption{\label{Tab_degrees} Summary of results for the five major currencies.
From left, the first column reports the $Z$ validation threshold. THe following reports the currency tickers. Then the following three columns report the degree in the valid cross correlation networks for prices, positive sentiment and negative sentiment. 
The following two columns report respectively the impacting and impacted degree for the positive sentiment - price valid correlation network. 
Finally, the last four columns report degrees in the valid transfer entropy network. 
 }
\end{table}

\subsection{ Price-Sentiment  transfer entropy causality network }
In order to quantify causal relations between sentiment and price in the cryptocurrency market, we computed  non parametric transfer entropy between  log variation of positive sentiment volume and log variations of price and vice versa.
These are two $1944\times1944$  asymmetric matrices  representing bipartite directed networks.

The diagonals of these matrices report respectively the causal influence of sentiment over price and the causal influence  of price over sentiment for each currency. 
As for the correlations we retain only the valid entries (over 40 common observations and $Z>3$).
We observed that the overall information flow (difference between the transfer entropy between sentiment to price and price to sentiment) is positive indicating for each currency that more information is transferred from past price to future sentiment than the contrary. 
However, only about 2\% of currencies have valid causality relations with 19 currencies having stronger causal influence of price over sentiment and, conversely, other 11 currencies with stronger causal influence of sentiment over price.
\CB{Interestingly, none of the  five major  currencies has valid internal price-sentiment causality in either directions. }

\begin{figure}
\centering
\includegraphics[width=.45\linewidth]{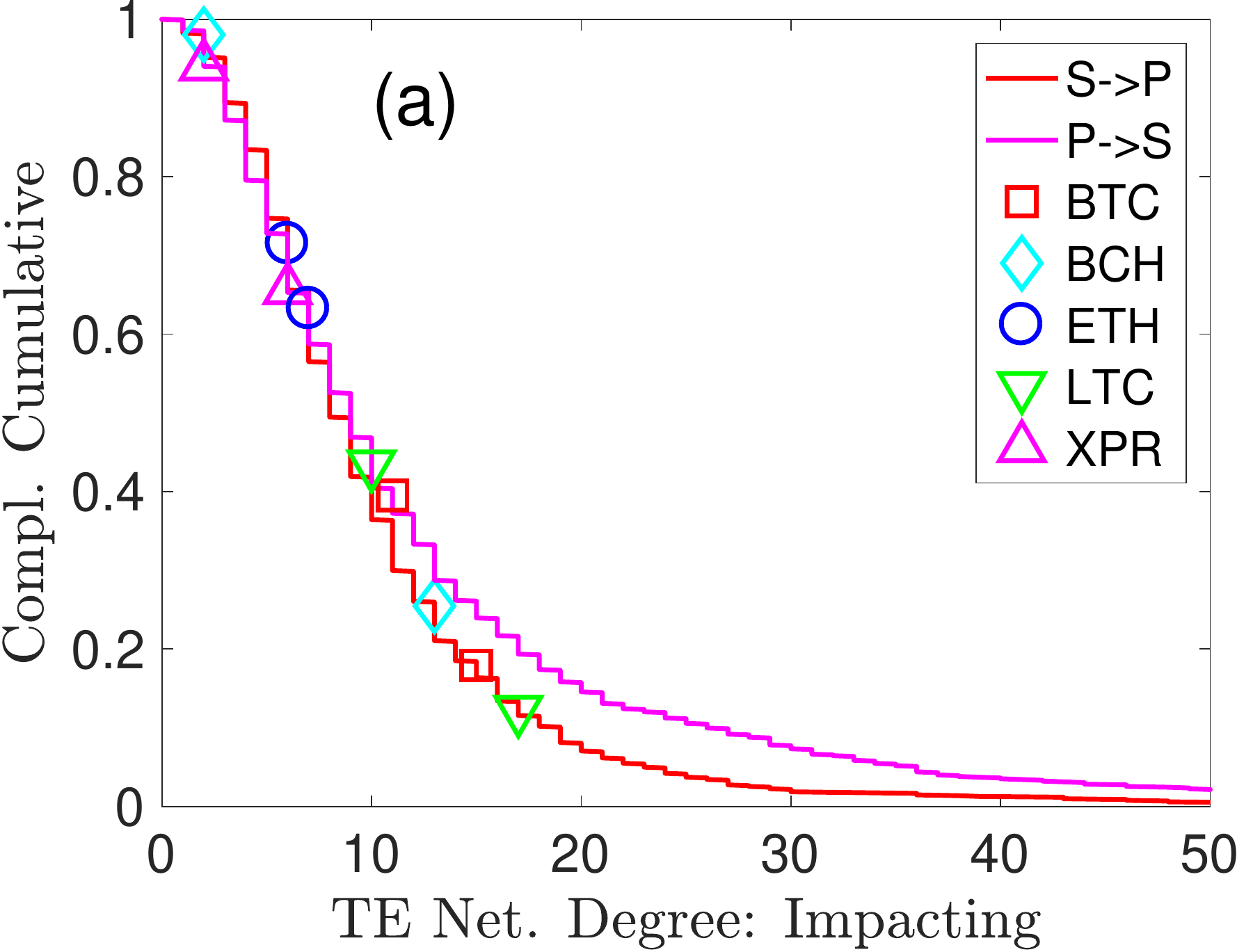}
\includegraphics[width=.45\linewidth]{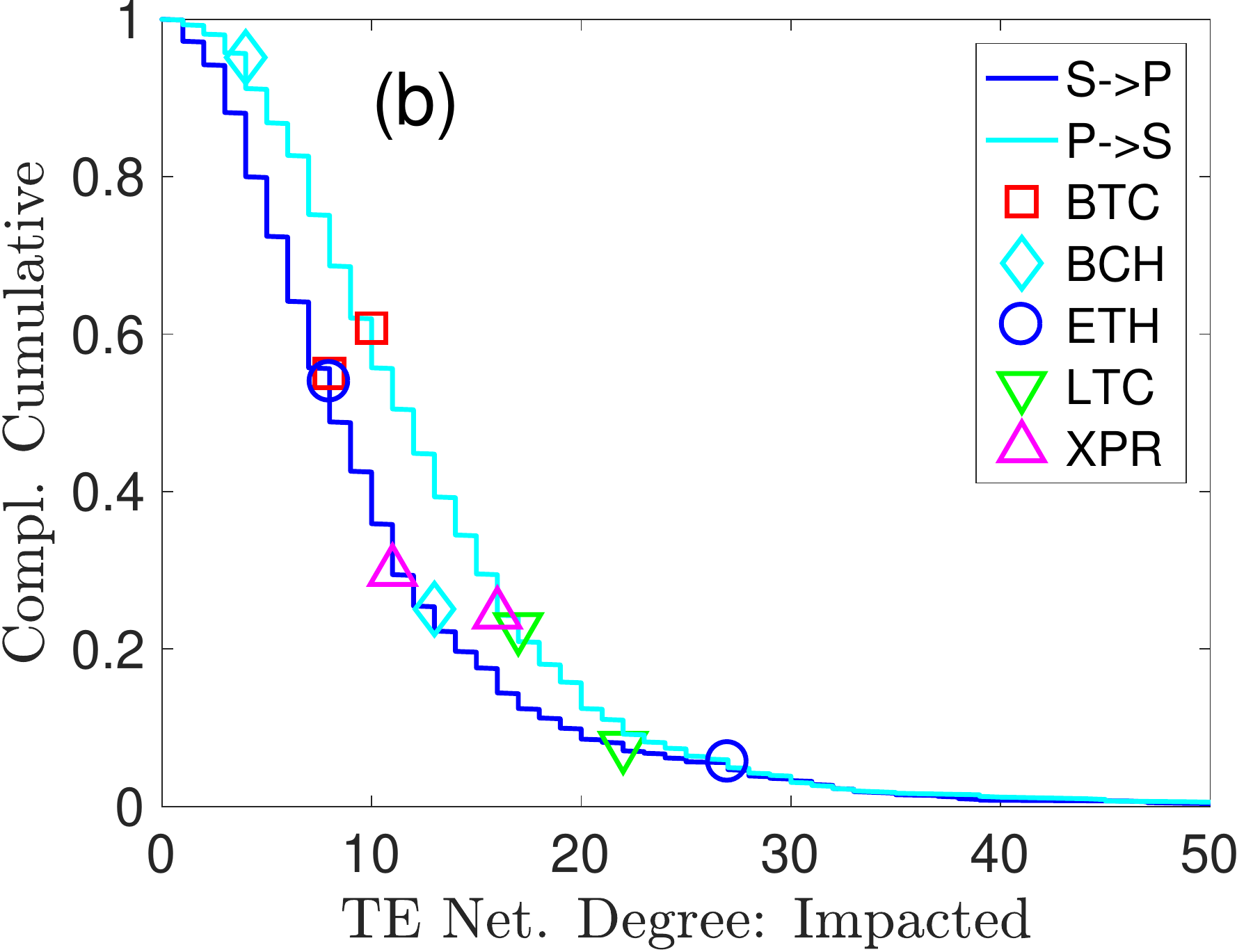}
\caption{Complementary cumulative  degree distributions  for the validated transfer entropy network.
(a) 'impacting' distribution: number of other currencies influenced by a given currency.
(b) 'impacted' distribution: number of other currencies influencing a given currency.
The plots report both the validated transfer entropy network for prices causing sentiment and the network for sentiment causing prices.
}
\label{Fig_TEnet}
\end{figure}

The off-diagonal elements estimate the causal influence between sentiment in currency $i$ on price of currency $j$ and, conversely, the causal influence between price in currency $i$ on sentiment of currency $j$.
These matrices are sparse  with  only about 0.3\%  valid entries (about 10,000 causality links).
Here we observed that the overall information flow is in the direction sentiment to price  indicating that the past sentiment of other currencies influences the future price of a given currency more than the effect of past prices over future sentiment. 
Conversely if we count the number of validated causality links we observe 13,179 causality links for prices causing sentiment and instead 10,352 for sentiment causing prices. 
The price causing sentiment network has average degree 6.8 and it has one giant component with 1023 elements.
Similarly, the sentiment causing price network has average degree 5.3 and one giant component with 1018 elements.
The degree distributions of the causality networks are shown in Fig.\ref{Fig_TEnet}. 
As in the previous case, we report two distributions: the `impacting'  and the `impacted',
the first being the number of all other currencies that act a valid causality over  a given currency, the latter being  the number of all other currencies that react with valid causality from a given currency.
These two degrees are computed for both the Price causing Sentiment and the Sentiment causing Price networks.
We observe that the five major currencies are spread in a central region of the ranking with respect to the other currencies, with Bitcoin sentiment being among the most impactful on other currency prices but with Bitcoin price being the least impacted by other currency sentiment. 

Summary of the results for the major currencies is reported in the last three columns of Table \ref{Tab_degrees}. 
We indeed can see that BTC positive sentiment is causing prices in 15 other currencies whereas only 8 other currencies sentiment are causing BTC price. We also note that ETH positive sentiment is the most impacted by other currencies prices and LTC price is caused by the largest number of other currencies positive sentiment. Finally, BCH causality is driven by sentiment much more than by prices.

We analyzed whether the relative position of a currency in the price network has an effect on the relation between this currency and sentiment. To this end we looked at the top 25\% most central currencies in the price cross correlation network in terms of weighted betweenness centrality. We then computed the transfer entropies of price causing sentiment and sentiment causing prices for these currencies and compared the number of causal relations with the ones for the bottom 25\% most peripheral currencies in the price cross correlation network. Results show that central currencies have ten times more causality links than the peripheral counterparts. Indeed, the top 20\% central currencies account already for 50\% of total causality links.  Intriguingly,  the signal is larger for sentiment causing prices than for prices causing sentiment.

\subsection{ Network significance from the comparison between price and sentiment networks  }
We have analyzed very noisy data that follow non-normal distribution and we tested millions of relations between variables. Spurious dependency and causality relations are certainly present. What we must test is if the structural properties we unveiled are real to the system or only spurious consequences of noise and randomness. 
To this purpose we first tested different levels of validation from $Z>2$ to $Z>6$ verifying that the results are consistent and persistent for different validation thresholds.
Some of these results for $Z>6$ are reported in the  bottom part of table \ref{Tab_degrees}.
Note that, within normal statistics assumptions $Z>6$, would correspond to p-values below $10^{-9}$ and nonetheless we still retrieve some of the  results previously reported especially for the price cross correlation network that is still highly connected.
However, we  also observed that at this threshold the transfer entropy network does not have any longer a giant component with the larger cluster having only 36 elements and average degree being 0.1. 
Overall, this analysis at large $Z$ thresholds gives us some confidence but still provides us with inconclusive answers about the significance of the results, indeed the non-normality of the statistics can strongly affect the corresponding statistics of the Z-score with sizeable likelihood of spurious results even at this threshold levels.

\CB{We therefore decided to adopt a different approach and, instead of trying to statistically validate each network, we cross-validate results  by comparing  metrics from networks build from unrelated signals, namely, the price, the positive and the negative sentiment. We argue that if, for instance, the network from sentiment correlations has significantly similar properties with the network from price correlations it is highly unlikely that the two represent random spurious correlations.}
We therefore compared the degree centrality (degree of each vertex) of the varius networks at different validation thresholds. We used Superman correlation for the quantification of the similarity between these measures. \CB{Results are reported in Tab.\ref{Tab_Net_Comp} where we can see that there are large and statistically significant correlations (t-test p-values smaller than $10^{-45}$) between all networks analyzed in this paper at all levels of  validation thresholding from $Z>3$ to $Z>6$.}
We note that similarities between the \CB{correlation  networks tend to increase with thresholding value up to $Z^* = 4$  and then decrease afterwards. Whereas the similarities with the combined Transfer Entropies network has maxima at $Z^*=3$.
The similarity increase with $Z^*$ in the correlation networks is consequence of the reduction in the noise and the decrease is instead the consequence of the reduction in statistics.
In the table, we did not include results from the sentiment-price network to avoid confusion and also because they are less significant given that the network is already build from the two signals.  Yet, results are well in line with the one reported in Table \ref{Tab_Net_Comp} with correlations ranging between 90\% to 45\%.}

\begin{table}
\begin{center}
\begin{tabular}{ |c|c|c|c|c|c|c| } 
\hline 
$Z^*$&P-pS &P-nS&pS-nS&TESP-P&TESP-pS&TESP-nS \\ \hline
3       & 0.42 &0.43&0.58   & 0.80 & 0.69  &0.57   \\ \hline
4       & 0.63 &0.54&0.72   & 0.75 & 0.61  &0.49   \\ \hline
5       & 0.58 &0.50&0.68   & 0.61 & 0.51  &0.41   \\ \hline
6       & 0.49 &0.46&0.60   & 0.45 & 0.38  &0.32   \\ \hline 
\end{tabular}
\end{center}
\caption{\label{Tab_Net_Comp}
\CB{
Spearman correlations between degree centralities in the dependency and causality networks from prices and sentiment signals.
Rows are different levels of validation threshold with $Z>Z^*$.
Columns are Spearman correlation coefficients between degree centrality measures of different networks.   
Specifically: P is the symbol for the prices network from Kendall correlations; pS is the symbol for the positive sentiment network from Kendall correlations; nS is the symbol for the negative sentiment network from Kendall correlations; TESP is the symbol for the combined Transfer Entropy causality networks between prices and sentiment.
The combined transfer entropy degree centrality is the sum of all edges incoming in and outgoing from each vertex in the transfer entropy networks.  
Statistical validation of the correlation values (t-test) give  p-values below $10{-45}$ for all these correlations. }
 }
\end{table}

%\begin{table}
%\begin{center}
%\begin{tabular}{ |c|c|c|c|c|c|c| } 
%\hline
%$Z^*$&P-pS&P-nS&pS-nS&TE-P&TE-pS&TE-nS \\ \hline
%3       & 0.42 &0.43&0.58   & 0.72 & 0.46  &0.43   \\ \hline
%4       & 0.51 &0.47&0.66   & 0.70 & 0.52  &0.47   \\ \hline
%5       & 0.54 &0.47&0.68   & 0.57 & 0.50  &0.41   \\ \hline
%6       & 0.46 &0.44&0.60   & 0.42 & 0.38  &0.32   \\ \hline
%\end{tabular}
%\end{center}
%\caption{\label{Tab_Net_Comp}
%Spearman correlations between degree centralities in the dependency and causality networks from prices and sentiment signals.
%Rows are different level of validation threshold with $Z>Z^*$.
%Columns are Spearman correlation coefficients between log-variations of: P-pS  prices and  positive sentiment; P-nS prices and negative sentiment; pS-nS positive sentiment and negative sentiment; TE-P combined transfer entropy prices-positive-sentiment  and prices; TE-pS combined transfer entropy prices-positive-sentiment  and negative sentiment.
%The combined transfer entropy degree centrality is the sum of all edges incoming in and outgoing from each vertex in the transfer entropy networks.  
%\CB{Statistical validation of the correlation values (t-test) give zero p-values (within numerical precision) for all these correlations. }
% }
%\end{table}

%CORRELATIONS (LARGE) BETWEEN THE THREE LAYER CENTRALITY AND CLOSENESS MEASURES -> PROOF THAT SENTIMET STRUCTURE IS NOT RANDOM

\section{Discussion} \label{S.discussion}
Our first and most important comment concerning this work is that data are very noisy. Price data have a slightly stronger signal than sentiment ones but in both cases noise is predominant. Nonetheless, we observe the presence of a structural organization both in the correlations and in the transfer entropy \CB{and we demonstrated that such a structure is not random.}
%Overall we observe that currency price changes are correlated to each other with predominantly positive correlations. 

Concerning the correlation analytics we have seen that price-price dependency have larger correlations but sentiment-sentiment and also sentiment-prices show valid and positive correlations.
Not surprising, we observed that Bitcoin and the other four major currencies have strong dependency ties with the prices of a vast number of other currencies. More surprisingly, we observed that, in contrast, in the sentiment dependency network these major cryptocurrencies are not highly connected. 
This is also reflected in the closeness and centrality measures that see all major currencies in non-central positions in the network with exception only for the closeness measure for the price network.
The sentiment-price correlation network also reflects mainly positive dependencies with major currencies  having only average or  just slightly above average degrees with exception for the dependency between Bitcoin sentiment and other currency prices that reveal instead very strong dependency connections. 

The transfer entropy has a lower fraction of valid links. This is mainly due to the fact that this measure requires the estimate of a probability distribution between three variables which is hard to estimate well with the short time-series we have.
Nonetheless, we observe a sizable fraction of valid causality links with most information flowing from prices to sentiment for each currency but instead from sentiment to price when the cross-effect of a currency on another is considered. 
Interestingly, in terms of number of valid links we observe   a larger number of causality links for prices causing sentiment  than for sentiment causing prices.
\CB{This indicates that causality of sentiment over price carries a larger amount of information but also a larger amount of noise and therefore it is validated  only at higher transfer entropy values.}
%Consistently with the correlation data we observe that the  five major  currencies have a relatively particular 

\CB{The comparison between causality of the central nodes in the prices network with respect the peripheral ones for what concerns the effect of sentiment over prices and prices over sentiment shows that currencies that are central to the systems in term of price behavior are also the ones that most strongly influence the sentiment in the whole system. This is an interesting finding also in the light of the results in \cite{pozzi2013spread} that uncovered the great difference between central and peripheral vertices in terms of investment performances and risk.
Note that the centre of the prices correlation network contains the five major currencies, however they are not the main responsible for the causality effect.}
 
We already stressed  that in this work we have investigated only valid dependency and causality links giving  us some confidence that weak noisy links are removed. However, statistics is not normal and in our system we have almost four million  possible relations between variables and some might turn out to be validated just as the effect of random fluctuations. 
We argued that the proof that, overall, results are robust and not reporting just incidental spurious relations must be searched in the similarity of metrics of networks extracted with different methodologies (Kendall correlations or Transfer Entropy) from different signals (prices or sentiment). In this respect the strong correlations reported in Tab.\ref{Tab_Net_Comp} are a good indication that these systems have a consistent structural organization with prices and sentiments influencing each-other in a significant way.

\section{Conclusions} \label{S.conclusion}
This study demonstrates that the current cryptocurrency market has a complex structure.
\CB{ Major, highly capitalized cryptocurrencies and minor little capitalized ones are interlocked into this complex structure with major currencies playing central roles only for the price dependency network. }
Sentiment and prices are interconnected and they show both dependency and causality mainly between different currencies. 

\CB{Social sentiment plays a very important role in this market with Bitcoin sentiment correlating with other currencies prices even more than with its own price and  with validated causal measures showing that sentiment is more influential on price than the contrary.}

\CB{An unexpected outcome of this research is  that minor low-capitalised currencies are playing a very important role in moving the market sentiment and consequently are significantly affecting prices also of the highly capitalised currencies. This is a fundamental difference from traditional markets where the driving economic factors are typically reflected into the dependency and causality structure \cite{aste2010correlation,song2012hierarchical,musmeci2014clustering}. The fact that economically irrelevant variables can  have influence on the whole structure of the system is however a typical feature of complex systems where the system cannot be understood from the analysis of its parts in isolation \cite{aste2010introduction}. This indicates that the study of cryptocurrencies and more generally of the digital economy require the development of tools beyond traditional approaches with use of instruments from the  science of complex systems.

Cryptocurrencies are increasingly traded and are becoming part of mainstream investment choices. From a risk-management and investment perspectives the present investigation unveil that the overall market dynamics is dominated by noise, large volatility and large failure rates. This is therefore a highly risky domain where most of the traditional risk management  and asset allocation instruments are likely to be ineffective. Complex system science \cite{aste2010introduction} can guide us into the development of new tools  for modelling, managing risk and design investment strategies for these markets and the new digital economy. }
%Further work is needed to device robust tools to operate on this novel market. }

%In this paper we wanted to provide an overall study of the market structure including all available currencies. A more specialized focus on the dependency and causality structure of the highly capitalized currencies only will be the object of a future work.
\CB{This paper is a first attempt to explore the very vast and intricate field of cryptocurrency market. Our efforts have been mostly dedicated to perform a statistically rigorous investigation of the whole market  by using innovative tools such as network measures, non-linear quantification of dependency and causality and non-parametric validation techniques. The results are robust despite the very challenging task to infer, from short time-series, non-linear interrelations in a very large multivariate system.

These are extremely dynamical systems that change continuously. Our analysis is limited to a short period of time and the system has already changed  between the time when the system was analysed and the publication of this paper. This is an unavoidable reality in these system and the contribution of this paper is not primarily about the actual specific properties of the cryptocurrencies market during the period investigated but some general facts, such as the influence of minor currencies on the whole system, that are likely to remain in the future and to be also characteristic of other systems. Further, an important contribution of this paper is the introduction of a set of rigorous innovative  methodologies for the study of  systems composed of a very large set of variables with non-liner interactions and with small numbers of available observations. This is a very general challenge common to most socio-economic and complex systems where the methods introduced with this paper can be conveniently adopted in the future.

Much more must be done in future. For instance, in the study of the interactions between prices and sentiment we neglected, for simplicity, the negative sentiment. It is however clear that this plays a very important role which appears to be not trivially related to the positive one. We also made many choices, starting from the $Z$ statistics validation threshold or the use of log-variation of sentiment volumes or the choice of considering all currencies and not only the few with relevant market share. Different choices produce different results. In our investigations we verified that the overall reported results  are robust and these are retrieved similarly by adopting different choices. However, a more extensive and systematic study is necessary.}

\section*{Acknowledgmets}
Many thanks to the PsychSignal team for providing the sentiment data. We wish to thank Yuqing Long for his careful data preparation. Also thank to Zac Keskin for investigating transfer entropy in these systems within a parallel work that helped clarifying several issues in this one as well. Many thanks to the UCL-FCA group for help with discussions and proof reading. Finally we are grateful to  EPSRC for funding the BARAC project (EP/P031730/1) and to EU for funding the FinTech project (H2020-ICT-2018-2 825215).

%\bibliography{BiblioCrypto}

\begin{thebibliography}{10}

\bibitem{marketcap}
https://coinmarketcap.com/.

\bibitem{gkillas2018extreme}
Konstantinos Gkillas, Stelios Bekiros, and Costas Siriopoulos.
\newblock Extreme correlation in cryptocurrency markets.
\newblock 2018.

\bibitem{szetela2016dependency}
Beata Szetela, Grzegorz Mentel, and Stanis{\l}aw Gedek.
\newblock Dependency analysis between bitcoin and selected global currencies.
\newblock {\em Dynamic Econometric Models}, 16(1):133--144, 2016.

\bibitem{kim2016predicting}
Young~Bin Kim, Jun~Gi Kim, Wook Kim, Jae~Ho Im, Tae~Hyeong Kim, Shin~Jin Kang,
  and Chang~Hun Kim.
\newblock Predicting fluctuations in cryptocurrency transactions based on user
  comments and replies.
\newblock {\em PloS one}, 11(8):e0161197, 2016.

\bibitem{kaminski2014nowcasting}
Jermain Kaminski.
\newblock Nowcasting the bitcoin market with twitter signals.
\newblock {\em arXiv preprint arXiv:1406.7577}, 2014.

\bibitem{stocktwits}
https://stocktwits.com/.

\bibitem{aste2010correlation}
Tomaso Aste, W~Shaw, and Tiziana Di~Matteo.
\newblock Correlation structure and dynamics in volatile markets.
\newblock {\em New Journal of Physics}, 12(8):085009, 2010.

\bibitem{song2012hierarchical}
Won-Min Song, T~Di~Matteo, and Tomaso Aste.
\newblock Hierarchical information clustering by means of topologically
  embedded graphs.
\newblock {\em PLoS One}, 7(3):e31929, 2012.

\bibitem{musmeci2014clustering}
Nicol{\'o} Musmeci, Tomaso Aste, and Tiziana di~Matteo.
\newblock Clustering and hierarchy of financial markets data: advantages of the
  dbht.
\newblock {\em CoRR}, 2014.

\bibitem{cryptocompare}
https://www.cryptocompare.com/.

\bibitem{PsychSignal}
https://www.psychsignal.com, October 2015.

\bibitem{kendall1938new}
Maurice~G Kendall.
\newblock A new measure of rank correlation.
\newblock {\em Biometrika}, 30(1/2):81--93, 1938.

\bibitem{schreiber2000measuring}
Thomas Schreiber.
\newblock Measuring information transfer.
\newblock {\em Physical review letters}, 85(2):461, 2000.

\bibitem{tungsong2017relation}
Sachapon Tungsong, Fabio Caccioli, and Tomaso Aste.
\newblock Relation between regional uncertainty spillovers in the global
  banking system.
\newblock {\em arXiv preprint arXiv:1702.05944}, 2017.

\bibitem{campbell1997econometrics}
John~Y Campbell, Andrew~W Lo, Archie~Craig MacKinlay, et~al.
\newblock {\em The econometrics of financial markets}, volume~2.
\newblock princeton University press Princeton, NJ, 1997.

\bibitem{pozzi2008dynamical}
F~Pozzi, T~Aste, G~Rotundo, and T~Di~Matteo.
\newblock Dynamical correlations in financial systems [6802-54].
\newblock In {\em PROCEEDINGS-SPIE THE INTERNATIONAL SOCIETY FOR OPTICAL
  ENGINEERING}, volume 6802, page 6802. International Society for Optical
  Engineering; 1999, 2008.

\bibitem{pozzi2008centrality}
Francesco Pozzi, Tiziana Di~Matteo, and Tomaso Aste.
\newblock Centrality and peripherality in filtered graphs from dynamical
  financial correlations.
\newblock {\em Advances in Complex Systems}, 11(06):927--950, 2008.

\bibitem{wilks1932certain}
Samuel~S Wilks.
\newblock Certain generalizations in the analysis of variance.
\newblock {\em Biometrika}, pages 471--494, 1932.

\bibitem{kendall1946advanced}
Maurice~George Kendall et~al.
\newblock The advanced theory of statistics.
\newblock {\em The advanced theory of statistics.}, (2nd Ed), 1946.

\bibitem{newman2008mathematics}
Mark~EJ Newman.
\newblock The mathematics of networks.
\newblock {\em The new palgrave encyclopedia of economics}, 2(2008):1--12,
  2008.

\bibitem{granger:econ}
Clive Granger.
\newblock Investigating causal relations by econometric models and
  cross-spectral methods.
\newblock {\em Econometrica}, 37(3):424--38, 1969.

\bibitem{granger1980testing}
Clive~WJ Granger.
\newblock Testing for causality: a personal viewpoint.
\newblock {\em Journal of Economic Dynamics and control}, 2:329--352, 1980.

\bibitem{PhysRevLett.103.238701}
Lionel Barnett, Adam~B. Barrett, and Anil~K. Seth.
\newblock Granger causality and transfer entropy are equivalent for gaussian
  variables.
\newblock {\em Phys. Rev. Lett.}, 103:238701, Dec 2009.

\bibitem{pozzi2013spread}
Francesco Pozzi, Tiziana Di~Matteo, and Tomaso Aste.
\newblock Spread of risk across financial markets: better to invest in the
  peripheries.
\newblock {\em Scientific reports}, 3, 2013.

\bibitem{aste2010introduction}
Tomaso Aste and Tiziana Di~Matteo.
\newblock Introduction to complex and econophysics systems: A navigation map.
\newblock In {\em Complex physical, biophysical and econophysical systems},
  pages 1--35. 2010.

\end{thebibliography}
%\bibliographystyle{unsrt} 

\end{document}